\documentstyle[11pt]{article}
\input epsf
\makeatletter
\renewcommand\theequation{\thesection.\arabic{equation}}
\@addtoreset{equation}{section}
\makeatother

\begin{document}
\thispagestyle{empty}
\baselineskip 20pt
\rightline{CU-TP-890}
\rightline{CLNS-98/1556}
\rightline{{\tt hep-th}/9804174}

\def\CN{{\cal N}}
\def\tr{{\rm tr}\,}
\newcommand{\beq}{\begin{equation}}
\newcommand{\eeq}{\end{equation}}
\newcommand{\beqn}{\begin{eqnarray}}
\newcommand{\eeqn}{\end{eqnarray}}
\newcommand{\bde}{{\bf e}}
\newcommand{\balpha}{{\mbox{\boldmath $\alpha$}}}
\newcommand{\bsalpha}{{\mbox{\boldmath $\scriptstyle\alpha$}}}
\newcommand{\bbeta}{{\mbox{\boldmath $\beta$}}}
\newcommand{\bsbeta}{{\mbox{\boldmath $\scriptstyle\beta$}}}
\newcommand{\blambda}{{\mbox{\boldmath $\lambda$}}}
\newcommand{\bslambda}{{\mbox{\boldmath $\scriptstyle\lambda$}}}
\newcommand{\ggg}{{\boldmath \gamma}}
\newcommand{\ddd}{{\boldmath \delta}}
\newcommand{\mmm}{{\boldmath \mu}}
\newcommand{\nnn}{{\boldmath \nu}}

\newcommand{\bra}[1]{\langle {#1}|}
\newcommand{\ket}[1]{|{#1}\rangle}
\newcommand{\sn}{{\rm sn}}
\newcommand{\cn}{{\rm cn}}
\newcommand{\dn}{{\rm dn}}
\newcommand{\diag}{{\rm diag}}

\vskip 1cm
\centerline{\LARGE\bf Dyons in $N=4$ Supersymmetric Theories} 
\centerline{\LARGE\bf and Three-Pronged Strings}
\vskip 1.2cm
\centerline{\large\it Kimyeong Lee$\;^{a}$\footnote{electronic mail:
klee@phys.columbia.edu} and Piljin Yi$\;^b$\footnote{electronic mail:
piljin@mail.lns.cornell.edu}} 
\vskip 2mm
\centerline{$^a$Physics Department, Columbia University, New York, NY
10027}
\vskip 1mm
\centerline{$^b$F.R. Newman Laboratory of Nuclear Studies, Cornell
University, Ithaca, NY 14853}
\vskip 1.5cm
\centerline{\bf ABSTRACT}
\vskip 2mm
\begin{quote}
We construct and explore BPS states that preserve $1/4$ of
supersymmetry in $N=4$ Yang-Mills theories. Such states are also
realized as three-pronged strings ending on D3-branes. We correct the
electric part of the BPS equation and relate  its
solutions to the unbroken abelian gauge
group generators. Generic $1/4$-BPS solitons are not spherically
symmetric, but consist of two or more dyonic components held apart by
a delicate balance between static electromagnetic  force and scalar
Higgs force. The instability previously found in three-pronged string
configurations is due to excessive repulsion by one of these static
forces. We also present an alternate construction of these $1/4$-BPS
states from quantum excitations around  a magnetic monopole, and
build up the supermultiplet for arbitrary (quantized) electric charge. 
The degeneracy and the highest spin of the supermultiplet 
increase linearly with a relative electric charge. We conclude with 
comments.
\end{quote}

%\pacs{14.80.Hv,11.27.+d,14.40.-n}

\newpage
\section{Introduction}

Among supersymmetric theories that are known to admit a strong-weak
coupling duality, $N=4$ $D=4$ supersymmetric Yang-Mills field
theories are perhaps the easiest and most straightforward to study. In
its Coulomb phase, the solitonic spectra are scrutinized in great
detail, where a manifest strong-weak coupling duality was observed
among the charged BPS particles that break exactly half of
supersymmetry. This includes the usual BPS magnetic monopoles and
standard dyonic excitations thereof whose electric charges are
proportional to the magnetic charge.  These BPS monopoles and dyons
break half of $N=4$ supersymmetry, and duality predicts that they
are all in the $N=4$ vector multiplet with the maximum spin 1, a
short multiplet of degeneracy $2^4=16$.

There are, however, other kinds of supersymmetric states which break
$3/4$ of supersymmetry. Such states would come in an intermediate
multiplet which contains spin $3/2$ or higher.  It is only very
recently that their properties are explored. Most notable is a work by
O. Bergman~\cite{oren} who constructed such dyons as three-pronged
strings that end on three parallel D3-branes. Here, we recapitulate
this construction.

Recall that $N=4$ $D=4$ $U(n)=SU(n)\times U(1)$ Yang-Mills theory is
a world-volume theory of $n$ parallel D3-branes~\cite{witten}.  The
Coulomb phase of the $U(n)\rightarrow U(1)^n$ theory is parameterized
by six adjoint Higgs expectations, whose $6n$ eigenvalues encode the
positions of the $n$ D3-branes in the internal part $R^6$ of the
spacetime $R^6\times R^{3+1}$.  One special feature of the D3-brane is
that it is self-dual under the $SL(2,Z)$ U-duality of the type IIB
string theory. As far as the low energy world-volume physics goes, a
practical consequence of this is that any $(q,g)$-string may end on
the D3-brane. Here $q$ and $g$ are the charges with respect to the two
antisymmetric tensor fields $ B_{\mu\nu}$ and ${\tilde B}_{\mu\nu}$ that
lives, respectively,  in the NS-NS
sector  and in the Ramond-Ramond sector of the type IIB theory. With
respect to the unbroken $U(1)$ 
associated with the D3-brane where a $(q,g)$ string end, then, such an
end-point appears as a particle of  $q$ electric and $g$ magnetic 
charges. The familiar BPS $(q,g)$ dyons of $SU(n)$ theory corresponds
to a straight $(q,g)$ string segment that connects a pair of
D3-branes.

A novelty comes from the fact that three-pronged strings are also in
the spectrum of string theory/M theory. They can be used to connect a
set of three D3-branes. The three segments that meet at a single
junction must have different $(q,g)$'s to preserve some
supersymmetry~\cite{prong1,prong2}, so the resulting BPS state has
its electric charge not proportional to its magnetic
charge. Typically, it will break $3/4$ of the $N=4$
supersymmetry.\footnote{Three-pronged strings can also generate 
BPS states in $N=2$ theories \cite{oren2}. In such cases, they actually break
only half of supersymmetry.} We will use the phrase ``$1/4$-BPS state'' to
distinguish from the usual BPS states that break only half of
supersymmetry. For instance, suppose that we have $SU(n)$ broken
down to $U(1)^{n-1}$. Pick a pair of roots $\balpha$ and $\bbeta$ with
$\balpha^2=\bbeta^2=1$ and $\balpha\cdot\bbeta=-1/2$.  A state of
magnetic charge $m\balpha+ m\bbeta$ and of electric charge $n\balpha$
would then be $1/4$-BPS.

Now the question is how these $1/4$-BPS states are realized on the
field theory side. One might be tempted to look for a spherically
symmetric soliton. In fact, very recently, a special class of
$1/4$-BPS states in $SU(3)$ theory was found in a spherically
symmetric ansatz~\cite{hasimoto}. In terms of roots, these BPS
configurations carry magnetic charge of $2\balpha+2\bbeta$. However,
as will become clear in later sections, the existence of these
solutions is quite accidental and fails to illuminate how the general
$1/4$-BPS dyons are constructed in the field theory language. One
severe problem is that if their electric charge is, say, of the form $q
\balpha$, the real number\footnote{Recall the electric charge is not
quantized in classical dyon solutions, unlike the magnetic charge
which is quantized topologically.} $q$ is determined uniquely by the
Higgs vev's. (In the spherically symmetric case of the total magnetic
charge, $\balpha+\bbeta$, for instance, $q$ has to vanish for all
vev's.) Because of this, at generic point of vacuum moduli space, BPS
configurations of properly quantized electric charge ($q={\rm
integer}$) cannot be realized as a spherically symmetric classical
soliton.

In general, we expect the BPS configurations to be of an elongated
shape. Roughly speaking, it will consist of a pair of dyonic cores
which are bound but separated by some distance $R$. This is due to a
delicate balance between static electromagnetic  force and scalar
Higgs force. (See section 3.)  Once we realize this, it is almost
obvious that the amount of electric charge has to depend on the
separation $R$ as well as Higgs vev's.  What one misses by insisting
the spherical symmetry is this extra parameter $R$. With this picture
in mind, it is now clear that a BPS configuration of given electric
and magnetic charges will have some definite length $R$ that
parameterizes the deviation from the spherical symmetry.

This begs for another question: what happens in the limit of $R\rightarrow 
\infty$? Since it is electromagnetic and Higgs interaction that separates 
the two dyonic cores, a change in $R$ implies a change in electric
charge.  At $R\rightarrow \infty$, the electric charge of the
$1/4$-BPS state reaches a limiting value. In all cases we consider,
the charge will actually reach its maximum possible value.  Trying to
put an even larger electric charge will result in an instability and
cause the two cores to fly away from each other. The upper bound on
the electric charge can be also translated into a lower bound on a
linear combination of Higgs vev's with any given electric charge, in
which form the instability was found in the three-pronged string
configuration in Ref.~\cite{oren}.

The paper is organized as follows. In section 2, we derive the BPS
bound of the energy functional and write down the complete set of
equations that $1/4$-BPS dyons must satisfy. This corrects and
generalizes those in Ref.~\cite{fraser}. The magnetic part of the
equations are unaffected by the electric part. Given any purely
magnetic BPS solutions, the electric part is determined by solving a
single {\it four-dimensional} covariant Laplace equation of an adjoint
scalar. The existence of its solutions is tied to the existence of
$U(1)$ gauge zero-modes of the purely magnetic soliton, which
completes the existence proof of all the expected $1/4$-BPS dyonic
states corresponding to three-pronged strings. In section 3, we take
the specific example of $SU(3)$ broken to $U(1)^2$. The $1/4$-BPS
dyonic configuration of magnetic charge $\balpha+\bbeta$ is
constructed, from which we extract the relationship between Higgs
vev's, electric charges, and the separation length $R$. Important but
technical details involve ADHMN construction, which we put in the
appendices.  We digress in section 4, and compare the field theory
results to those from D-brane/three-pronged string picture.  The
instability bound is compared with that from the string construction,
and a perfect fit is found.

In section 5, we present an alternate construction of the $1/4$-BPS
dyons via exciting compactly supported eigenmodes around spherically
symmetric monopoles of magnetic charge $\balpha+\bbeta$. The correct
supermultiplet structure of $1/4$-BPS states are shown to be
reproduced, after a careful consideration of low energy eigenmodes.
The approximation, however, ignores some backreaction of the bosonic
background to excitation of these eigenmodes, which puts a stringent
criteria on the validity of the construction.  Because of this, in
particular, it is impossible to see the instability in this second
picture. In section 6, we use this construction to build up
the supermultiplet structure of dyons of arbitrary quantized electric
charge. Finally in section 7, we conclude with comments on 
unresolved issues.

\section{BPS Energy Bound and Equations}

Since the electric part of the BPS equations we found is different
from what is commonly known~\cite{fraser}, we will rederive the BPS
energy bound and equations from scratch. Also there are several
interesting new comments to be made about the BPS field
configurations. We start by considering the bosonic Lagrangian of the
$N=4$ supersymmetric Yang-Mills theories. With the gauge group
$SU(n)$ with hermitian generators $T^a$ in the $n$ dimensional
representation with the normalization $\tr T^a T^b= \delta^{ab}/2$, we
introduce the gauge field $A_\mu = A_\mu^a T^a$ and six Higgs fields
$\phi_I = \phi_I^a T^a$, $I=1,...,6$.  The bosonic Lagrangian density
is
\beq
{\cal L} = \tr \biggl\{ -\frac{1}{2} F_{\mu\nu}F^{\mu\nu} + D_\mu
\phi_I D^\mu \phi_I - \frac{1}{2} \sum_{I,J=1}^6 (-ie
[\phi_I,\phi_J])^2 \biggr\},
\eeq
where $D_\mu \phi_I = \partial_\mu \phi_I - ie [A_\mu,\phi_I]$.

\subsection{BPS bound}

The energy density is 
\beqn {\cal H} &=& \tr \biggl\{ (E_i)^2 + (B_i)^2 + (D_0\phi_I)^2+
(D_i\phi_I)^2 +\sum_{I<J}(-ie [\phi_I,\phi_J])^2 \biggr\} \nonumber \\
&=& \tr \biggl\{ (a_I E_i + b_I B_i - D_i\phi_I)^2 + (D_0\phi_I)^2
+\sum_{I<J}(-ie [\phi_I,\phi_J])^2 \biggr\} \nonumber \\ 
&+&2\;\tr\biggl\{  E_i D_i a\cdot\phi + B_i D_i b\cdot\phi \biggr\} ,
\eeqn
where $a_I,b_I$ are two arbitrary six-dimensional unit vectors
orthogonal to each other, $a\cdot\phi \equiv a_I\phi_I$ and $b\cdot\phi\equiv
b_I\phi_I$. The cross terms can be rewritten as,
\beqn
\tr B_i  D_i b\cdot \phi &=& \partial_i\,(\tr  b\cdot\phi B_i), \\
\tr E_i  D_i a\cdot \phi &=& \partial_i\, (\tr  a\cdot\phi E_i) 
-ie\, \tr \,(D_0\phi_I [a\cdot\phi,\phi_I]) , \label{addi}
\eeqn
where we used the Bianchi identity $D_i B_i=0$ and the Gauss law,
\beq
D_i E_i - ie[\phi_I, D_0\phi_I] = 0 .
\label{gauss}
\eeq
Denote collectively by $\zeta_I$, the  components of $\phi_I$ which are
orthogonal to both $a_I$ and $b_I$.  We split the energy 
density from the scalar fields into two parts;
\beq
 (D_0 a\cdot\phi)^2 +(D_0 b\cdot \phi)^2 +  (-ie [a\cdot\phi, b\cdot\phi])^2 ,
\eeq
and
\beq 
 (D_0 \zeta_I)^2 + (-ie [a\cdot\phi, \zeta_I])^2 + 
(-ie [b\cdot\phi, \zeta_J])^2 +\sum_{I<J} (-ie [\zeta_I, \zeta_J])^2,
\eeq
then complete the squares in the energy density as,
\beqn
{\cal H} &=& \tr\left\{( E_i -D_i a\cdot\phi)^2 +(B_i-D_ib\cdot\phi)^2 +
(D_0a\cdot\phi)^2 +(D_0 b\cdot\phi -ie [a\cdot\phi,b\cdot\phi])^2 \right\} 
\nonumber \\
         &+& \tr\left\{ (D_0 \zeta_I-ie [a\cdot\phi,\zeta_I])^2 +(D_i
          \zeta_I)^2+  (-ie [b\cdot\phi, \zeta_J])^2 + \sum_{I<J} 
(-ie [\zeta_I,   \zeta_J])^2 \right\} \nonumber \\
     &+& 2\partial_i\; \tr\left\{a\cdot \phi E_i + b\cdot\phi B_i \right\}. 
\eeqn
Every term except those in the last line is nonnegative, so the total 
energy is bounded by the contribution from the latter;
\beq
{\cal E}=\int d^3x \;{\cal H}  \; \ge \; {\rm Max}\;(a_I Q^E_I + b_I Q^M_I), 
\label{bps1}
\eeq
with
\beqn
Q_I^E &=& 2\int d^3x \;\partial_i\,( \tr \phi_I E_i ), \\
Q_I^M &=& 2 \int d^3x\; \partial_i\, ( \tr \phi_I B_i ). 
\eeqn
One most stringent bound must be found by varying $a_I$ and $b_I$ and
achieving the maximum. The quantities $Q_I^E$ and $Q_I^M$ can be evaluated
by converting to boundary integrals, and clearly depends on the asymptotics
only.

The expression $a_IQ^E_I+b_IQ^M_I$ is maximized only if the two unit 
vectors lie on the plane spanned by $Q^M_I$ and $Q^E_I$. Assuming this,
let $\alpha$ be the angle between $Q^M_I$ and $Q^E_I$, and $\theta$
the one between $b_I$ and $Q_I^M$. The extrema occur if and only if
\beq
\pm a_I Q^M_I = b_I Q^E_I,
\label{aqmbqe}
\eeq
which can be translated to an equivalent condition;
\beq
\tan \theta = \frac{\pm Q^E \cos\alpha}{Q^M\pm Q^E\sin\alpha}.
\eeq
$Q^M$ and $Q^E$ are the magnitude of vectors $Q^M_I$ and $Q^E_I$.
The two positive extrema are the
two central terms of $N=4$ supersymmetry algebra,,
\beq
Z_\pm = \sqrt{(Q^M)^2 +(Q^E)^2 \pm 2 Q^M Q^E \sin \alpha}.
\eeq
The true BPS bound for $N=4$ theory is then,
\beq
{\cal E} \ge {\rm Max}\,(Z_+, Z_-).
\eeq

\subsection{BPS Equations in Generic $N=4$ Vacua}

The BPS bound is saturated when every bulk term in the energy density
vanishes, from which we obtain total  eight sets of equations. The
first part is the most familiar; 
\beq B_i= D_i b\cdot\phi
\label{bbps}.  
\eeq 
This is the usual BPS equation that admits
magnetic monopole solutions.  Note that this magnetic equation can be
solved independently, regardless of remaining equations. The other BPS
equations influence only the choice of the unit vector $b_I$. This
fact is of crucial important when we construct the BPS solution
later. 

The second, electric part is made of several equations
\beqn
&& E_i= D_i a\cdot\phi \label{eeq}, \label{ebps} \\
&& D_0 a\cdot\phi=0 \label{0aphi},\\
&& D_0 b\cdot\phi=-ie[b\cdot\phi,a\cdot\phi] \label{0bphi}.
\eeqn
Using the latter two, we reduce the Gauss law (\ref{gauss}) to
\beq
D_i E_i = e^2\, [b\cdot\phi, [b\cdot\phi,a\cdot\phi] ]
+e^2[\zeta_I,[\zeta_I, a\cdot\phi]].
\label{gauss1}
\eeq
Combining this with Eq.~(\ref{eeq}) into a single second order linear
differential equation, we find that 
\beq
D_i D_i\, a\cdot \phi  = e^2\,[b\cdot\phi, [ b\cdot\phi,
a\cdot\phi]]+ e^2[\zeta_I,[\zeta_I,a\cdot\phi]].
\label{gauss15}
\eeq
which is a linear equation for $a\cdot\phi$ once $\zeta_I$'s are given.

So far we have not required that the spatial gauge field $A_i$ be 
time-independent. If we choose such a gauge, one sees easily that 
Eq.~(\ref{ebps}) is solved by
\beq
A_0 = -a\cdot\phi .
\eeq
In this gauge, $D_0\zeta_I-ie\,[a\cdot\phi,\zeta_I]=\partial_0\zeta_I=0$,
which requires $\zeta_I$ be time-independent. Other $\zeta_I$ equations
require them to be covariantly constant ($D_i\zeta_I=0$), commute with 
$b\cdot\phi$, and also commute among themselves: In the unitary gauge
where $b\cdot\phi$ is diagonal, the $\zeta$'s are all diagonal, constant,
and uniform, and also commute with the $A_i$'s. The latter condition 
implies that each $\zeta_I$ is proportional to the identity in each irreducible
block(s) spanned by nontrivial part of the configurations $A_i$ and 
$b\cdot\phi$.\footnote{If we were considering more general configurations with
many three-pronged strings connected to form a string web, this 
would translate to the requirement that the BPS string web is planar
in the internal space $R^6$.}  
Imagine that one think of the magnetic solution to Eq.~(\ref{bbps}) as
embedded along a subgroup of the original gauge group; the expectation value
$\zeta_I$'s must be invariant under such a subgroup.

Now Eq.~(\ref{gauss15}) is a zero-eigenvalue problem of a nonnegative
operator acting on $a\cdot\phi$ linearly.
Under the boundary condition that $a\cdot\phi(\infty)$ should commute 
with the asymptotics of $b\cdot\phi$ and $\zeta_I$, its solutions have 
nontrival behavior only in the said irreducible block(s). 
Thus $\zeta_I$ should also commute with $a\cdot\phi$. 
With such expectation value $\zeta_I$'s, Eq.~(\ref{gauss15}) reduces to,
\beq
D_i D_i\, a\cdot \phi  = e^2\,[b\cdot\phi, [ b\cdot\phi, a\cdot\phi]].
\label{gauss2}
\eeq
This is a four-dimensional covariant Laplacian for an adjoint scalar
field, provided that we identify $D_4\equiv- ie b\cdot\phi$. A more
restricted version of this equation, where one assumes 
$[b\cdot\phi,a\cdot\phi]=0$ as well, has appeared and been used in existing
literatures~\cite{fraser,hasimoto}. Thus, we find two sets of 
relevant BPS equations, given by Eq.~(\ref{bbps}) and 
(\ref{gauss2}), that must be solved to produce classical $1/4$-BPS
configurations.  (See Appendix E for a discussion about the
energy density of  BPS configurations.)

\subsection{Dyons and the Scalar BPS Equation}

The general configuration will have both magnetic and electric charges.
Along, say, $-z$ axis, the asymptotic behavior of the Higgs fields will
be
\beqn
& & b\cdot\phi \simeq b\cdot\phi(\infty) - 
\frac{{\bf g}\cdot {\bf H}}{4\pi r},
\label{bphib}\\
& & a\cdot\phi \simeq a\cdot\phi(\infty) - 
\frac{{\bf q}\cdot{\bf H}}{4\pi r}.
\label{aphib}\eeqn
The $n-1$ dimensional vectors ${\bf g}$ and ${\bf q}$ are the magnetic
and the electric charge, respectively, while ${\bf H}$ generates the 
Cartan subalgbra of $SU(n)$. 

We need to solve the first order equation (\ref{bbps}) and the second order 
equation (\ref{gauss2}). The first order equation is the well-understood 
BPS equation for monopoles~\cite{erick}. Let the vacuum
expectation values  of the Higgs be such that
\beq
b\cdot\phi(\infty) =  {\bf h}\cdot {\bf H} = {\rm diag}(h_1,h_2,..,h_n),
\eeq
where $\sum_a h_a=0$ and $h_1<h_2<...<h_n$.\footnote{These quantities
$h_i$ can be thought of as projected coordinate values of the $n$
D3-brane positions  along the $b_I$ direction. Thus, the gauge
symmetry could  be still broken even when some of $h_i$'s coincide.}
The magnetic charge of any BPS configuration 
should satisfy the topological quantization,
\beq
{\bf g}\cdot{\bf H}= \sum_{r=1}^{n-1}
 \frac{4\pi}{e}\,l_r\bbeta_r\cdot{\bf H} =
 \frac{2\pi}{e} \,{\diag}(-l_1,l_1-l_2,l_2-l_3,...,l_{n-1})
\eeq
with nonnegative integers $l_r$. One interprets such configurations as
being made of $n-1$ species of fundamental monopoles, where $l_r$ is
the number of the $r$-th fundamental monopole associated with the
simple root $\bbeta_r$. The conditions on the diagonal $\zeta_I$'s 
can be translated quite easily now. Generically, $\zeta_I$ must have 
vanishing inner products with all $\bbeta_r$ whenever $l_r\neq 0$. 
Only exception is when a consecutative chain of $\bbeta_r$ is such that
$l_s=\cdots=l_{s+t}$ and the corresponding monopoles are ``coincident.''
In that case, $\zeta_I$ must have a vanishing inner product with 
$\sum_{r=s}^{r=s+t}\bbeta_r$ but not necessarily with individual 
$\bbeta_r$, \dots, $\bbeta_{r+s}$.

The second-order BPS equation (\ref{gauss2}) is to be solved in the background
of  purely magnetic solutions to $B_i=D_i\,(b\cdot\phi)$. While we
will come back to actual solutions for specific examples next section,
it is important to note that the existence of the solution is already
well established. In fact, we know the exact number of linearly 
independent solutions. This is because any gauge zero mode of a BPS
monopole solution is automatically a solution to Eq.~(\ref{gauss2}).

Recall that the conventional way of finding zero-modes of BPS
monopoles is to perturb $B_i=D_i\Phi$ and impose the background gauge
$D_i\delta A_i =ie\,[\Phi,\delta\Phi]$~\cite{erick}. For a gauge
zero-mode, say, generated by a gauge function $\Lambda$, the
linearized BPS equations are always satisfied since both $B_i$ and
$D_i\Phi$ are gauge-covariant. Only the gauge-fixing condition is
nontrivial,
\beq
D_i\delta A_i =ie\,[\Phi,\delta\Phi] \qquad\Rightarrow\qquad
D_iD_i\Lambda=e^2[\Phi,[\Phi,\Lambda]].
\eeq
Inserting the solution to $B_i=D_i (b\cdot\phi)$ as the background
field, and replacing $\Lambda$ by $a\cdot\phi$, we realize that this
is identical to Eq.~(\ref{gauss2}).  The number of solutions to this
covariant Laplace equation, must equal the number of unbroken $U(1)$
generators that act nontrivially on the monopole solution. There must
be at least one and at most $n-1$.

Where is the electric charge located? 
When magnetic monopoles described by the first BPS
equation (\ref{bbps}) are well separated from each other, the field
configuration outside the core region would be purely abelian and so
cannot carry any electric charge. Each fundamental monopole
may carry only its own type of electric charge, that is, $\bbeta_r$
monopoles can carry only $\bbeta_r$ electric charge for any simple
roots $\bbeta_r$. One could say that generic 1/4-BPS configurations 
are made of classically bound (two or more) $1/2$-BPS dyons.

One might think that there is something odd about what we are doing
here.  After all, what we mean by $b\cdot\phi$ and $a\cdot\phi$ do
depend on what kind of electric and magnetic charges we have, yet we
seem to have fixed $b_I$ even before turning on the electric
charge. But what matters at the end of the day is that we get a set of
field configurations that solve all BPS equations simultaneously for
some $b_I$ and $a_I$. The BPS bound is a mini-max problem where one
tries to obtain a most stringent lower bound for all reasonably smooth
configurations. The simple fact that a configuration saturates a lower
bound, implies that the bound it saturates is actually the maximum
possible of all lower bounds. In section 3, we shall see how this is
realized in a concrete way.

\section{$1/4$-BPS Soliton in the $SU(3)$ Theory}

As an example, let us consider the $SU(3)$ gauge group.  Following the
strategy outlined in the previous section, we start with a purely
magnetic BPS configuration of a pair of distinct monopoles.  The
configuration must solve only the magnetic part of BPS equations, and
the scalar BPS equation will be solved in that background.

If we let $b\cdot\phi (\infty)$ be equal to ${\rm diag}\, (h_1,h_2,h_3) $ with 
$h_1<h_2<h_3$ and $h_1+h_2+h_3=0$, the two fundamental monopoles would
have magnetic charges,\footnote{Unless noted otherwise, we will 
suppress the electric coupling constant $e$ from now on.}
\begin{eqnarray}
4\pi\,\balpha\cdot {\bf H} &=& 2\pi\;{\rm diag}(-1,+1,0), \\
4\pi\,\bbeta \cdot{\bf H} &=& 2\pi \;{\rm diag}(0,-1,+1).
\end{eqnarray} 
We will label these monopoles by their charge vector in root space;
$\balpha$ and $\bbeta$. Throughout the rest of the paper, we will
consider 1/4-BPS configurations with magnetic charge of
$\balpha+\bbeta$. Accordingly, the asymptotic behavior of $b\cdot\phi$
would be
\beq 
b\cdot\phi\simeq {\rm diag}(h_1,h_2,h_3) -
\frac{(\balpha+\bbeta)\cdot{\bf H}}{r}.  
\eeq
{}From the work of E. Weinberg~\cite{erick}, we learn that the
separation between the two monopole cores is an arbitrary parameter,
which we denote by $R$.  $R$ uniquely determines $A_i$ and
$b\cdot\phi$ up to overall position, spatial orientation, and internal
gauge angles. The explicit form of the field configuration can be
obtained in principle from the ADHMN formalism~\cite{adhm,nahm}. The latter
is summarized in Appendices A and B. Recently, E. Weinberg  and
one of the authors (P.Y.) have found the explicit $A_i$ and $b\cdot\phi$
configuration for these two monopoles by exploring the Nahm's
formalism \cite{piljin}.

Now the difficult part is to solve the covariant Laplace equation;
\beq
D_i^2\Lambda=[b\cdot\phi, [b\cdot\phi, \Lambda]] .
\eeq
Once this is done, we simply take $a\cdot\phi$ to be a linear
combination of all possible solutions $\Lambda$.  We know, from the
arguments of previous section, there exist two linearly independent
solutions. We already know one such solution, since
$D_i^2(b\cdot\phi)=D_iB_i=0$ and $b\cdot\phi$ obviously commutes with
itself.  How do we find the other solution?  There have been several
works on the finding the solution of the covariant Laplacian of the
adjoint Higgs field around the instanton
background~\cite{osborn2}. This can be generalized to the magnetic
monopole background, which can be obtained as a limit of an instanton on
$R^3\times S^1$ with nontrivial Wilson 
loop~\cite{murray,caloron,kraan}. 
Appendix B and C provide the detailed discussion of the
solution for the covariant four-dimensional Laplacian.  Especially, a
single instanton in the $SU(3)$ case are made of three monopoles, two
of which correspond to two simple roots and one of which does to one
minimal negative root. This additional monopole solution depends on
the $x_4$ coordinate of $S^1$ and here we take the limit where this
additional monopole is taken to spatial infinity.

We will refer all detailed computation of the $SU(3)$ case to the
Appendix D. In this section, we will simply borrow the result and use
it for the study of (unquantized) 1/4-BPS configurations.  Combine the
Higgs expectation values to $\mu_2=h_3-h_2$ and $\mu_1=h_2-h_1$.  For
$SU(3)$ case, there are two independent solutions to the covariant
Laplace equations, since there are two unbroken $U(1)$'s acting on the
pair of monopole solutions. We will only need their asymptotic forms,
which can be read off from Eq.~(\ref{Lamb}).

As mentioned above, the first is proportional to the Higgs field $b\cdot\phi$ 
itself, whose asymptotics are
\beq
\Lambda_T \simeq {\rm diag}\left(h_1 + \frac{1}{2r}, h_2, h_3
-\frac{1}{2r}\right) ,
\label{lamb1}
\eeq
while the second is a bit more involved
\beq
\Lambda_R \simeq {\rm diag}\left(\mu_2 +\frac{p_1}{2r}, -(\mu_1+\mu_2)
+\frac{p_2-p_1}{2r}, \mu_1 - \frac{p_2}{2r} \right) .
\label{lamb2}
\eeq
The real numbers $p_1$ and $p_2$ are defined to be
\beqn
& & p_1 = \frac{\mu_1-\mu_2-2(\mu_1+2\mu_2)\mu_2
R}{\mu_1+\mu_2+2\mu_1\mu_2 R}, \nonumber \\
& & p_2 = \frac{\mu_1-\mu_2+2(2\mu_1+\mu_2)\mu_1
R}{\mu_1+\mu_2+2\mu_1\mu_2 R} .
\label{prel2}
\eeqn
$R$ is again the separation between the two monopoles, as naturally occurs
in the standard form of monopole moduli space metric or in the Nahm data.

The scalar field $a\cdot\phi$ and thus $A_0$ would be in general  
a linear combination of $\Lambda_T$ and $\Lambda_R$. Denote the respective
coefficients by $\xi$ and $\eta$;
\beqn
a\cdot\phi(\infty)&=&  \xi \, {\rm diag}(h_1,h_2,h_3) + \eta \,{\rm
diag}(\mu_2, -\mu_2-\mu_1, \mu_1) \nonumber \\
&=& \xi {\bf h\cdot H}+
2\eta\left(\mu_1\,\bbeta\cdot{\bf H} -\mu_2\,\balpha\cdot{\bf
H}\right).
\label{aphisu3}
\eeqn
The resulting electric charge is such that
\beq
{\bf q} = q_\alpha \balpha +q_\beta \bbeta,
\eeq
where
\beqn
&& q_\alpha = 4\pi (\xi + \eta p_1), \nonumber \\
&& q_\beta = 4\pi( \xi + \eta p_2 ).
\label{qaqb}
\eeqn
For any nonzero separation $R$, the electric charge is misaligned
against the magnetic charge unless $\eta=0$. For $R=0$, however,
electric charge is proportional to $\balpha+\bbeta$. For any $R$, it
is easy to double-check that the BPS configuration indeed saturate
the most stringent BPS bound. All one need to ensure is that the angle
$\theta$ between $Q^M_I$ and $b_I$ is unchanged as the electric charge
is turned on, which is in turn guaranteed as Eq.~(\ref{aqmbqe}) holds.
This is always true for the solution we obtained.

The resulting 1/4-BPS configuration is then composed of a pair of
distinct monopole separated by a distance $R$, and on top of which the
time-like gauge potential $A_0=-a\cdot\phi$ is turned on to carry the
additional electric charge whose relative value is completely
determined by $R$. The $\balpha$ monopole would carry $q_\alpha$
electric charge and the $\bbeta$ monopole would carry $q_\beta$
electric charge.  The relative electric charge $(q_\beta -q_\alpha)/2$
is the part of the electric charge orthogonal to the magnetic charge
and is given by
\beq
\Delta q =8\pi
\eta\; \frac{(\mu_1^2+\mu_1\mu_2+\mu_2^2)R}{\mu_1+\mu_2+2\mu_1\mu_2
R}.
\eeq 
This is responsible for the electromagnetic repulsion,
which must be balanced against the Higgs attraction.\footnote{It would
be interesting to derive this relative charge from the consideration of the
long range force law.}  Note that $\Delta
q$ is a monotonic function of $R$. In particular, $R=0$ implies that
$\Delta q=0$ as well.  When the two constituent monopoles form a
single spherically symmetric configuration, they can be $1/2$-BPS but
not $1/4$-BPS.

As $\Delta q$ increases, $R$ increases, and at some critical charge,
the separation diverges, $R\rightarrow \infty$. This of course signals
that the BPS configuration no longer exists as a single particle
state. Two solitonic cores are separated by arbitrarily large distance
once $\Delta q$ reaches its maximum possible value, 
\beq
\Delta q_{\rm
cr} = 4\pi\eta\, \frac{(\mu_1^2+\mu_1\mu_2+\mu_2^2)}{\mu_1\mu_2 },
\label{deltaq}
\eeq
at which point the instability sets in.
While we carried out the analysis with arbitrary electric charges,
it is simply a matter of putting particular values of $R$ if 
one wishes to extend the result to properly quantized dyons.

Before closing this section, we would like to clarify how a
spherically symmetric $1/4$-BPS dyon is possible for higher magnetic
charges. As we just saw, the only spherically symmetric solution with
magnetic charge corresponding to a root, say $\balpha+\bbeta$, is the
ones that break half of supersymmetry.  They cannot possess any
relative electric charge. However, when the magnetic charge is a
double, say $2\balpha+2\bbeta$, the analog of this $1/2$-BPS, purely
magnetic state is not spherically symmetric. The situation is
analogous to having a pair of identical $SU(2)$ monopoles as close to
each other as possible, if we consider the $SU(2)$ as embedded inside
$SU(3)$ along $\balpha+\bbeta$. We know from early works on $SU(2)$
monopoles that this configuration is cylindrically symmetric, and of
toroidal shape~\cite{ward}.  As we turn on relative electric charge and
thereby reduce the state to $1/4$-BPS, all four constituents, two
$\balpha$'s and two $\bbeta$'s, begin to move away from one another,
and eventually become independent. It is then conceivably that, at
some specific electric charge, the all four soliton cores are
separated just right so that they actually form a spherically
symmetric shape. The one solution found in Ref.~\cite{hasimoto}, is an
example of this phenomenon.

\section{Three-Pronged String and Instability}

Let us compare the above result against the string picture. For the
purpose of this section, we will pretend that string tension is not
quantized, since in the end the physics of instability can be
understood classically. Let us consider the specific configuration
with the $q$ fundamental strings and $g$ D-strings so that, in  the field
theoretic context, this is translated to a magnetic charge
$g(\balpha+\bbeta)$ and the electric charge $q_\alpha \balpha$. Take
$\xi=-p_2\eta$ so that $q_\beta=0$ of Eq.~(\ref{qaqb}), then the
dyonic solution of previous section acquires an electric charge along 
$\balpha$ only,
\beq
 {\bf q}=  4\pi\eta (p_1-p_2)\balpha.
\eeq
Let $q\equiv q_\alpha=4\pi\eta (p_1-p_2)$.

Let $X^I_{21}$ be the six-dimensional displacement between the first
and the second D3-branes, and similarly $X^I_{32}$ be the one between
the second and the third D3-branes. The projection along $b_I$ is
determined by the Higgs vev $b\cdot\phi(\infty)$; 
\beq
b_IX^I_{21}=h_2-h_1,\qquad b_IX^I_{32}=h_3-h_2, 
\eeq 
and similarly
$a\cdot\phi(\infty)$ of Eq.~(\ref{aphisu3}) determines the projection
along $a_I$.  The vectors $Q^E_I$ and $Q^M_I$ are then, 
\beqn 
&& Q_I^M
= g X^I_{31}= g\,(X^I_{32}+X^I_{21} ), \\ 
&& Q_I^E = q X^I_{21} ,
\eeqn
where
\beqn
&& X^I_{21}=(h_2-h_1)\, b_I-\eta(2\mu_2+\mu_1+p_2\mu_1)\,a_I,\\
&& X^I_{32}=(h_3-h_2)\,b_I+\eta (2\mu_1+\mu_2-p_2\mu_2)\,a_I.
\eeqn
A simple generalization of Bergman's calculation shows again
that the energy of the string configuration coincides with the field
theoretic one if we identify the string tension of $(q,g)$ string to
be $\sqrt{q^2+g^2}$ in the field theory unit. If we quantize the
system, $q$ becomes the number of the fundamental strings.  The same
consideration tells us that the angle $\omega$ between the $(0,g)$ string and 
the $(q,g)$ string as they meet at the junction is solely determined 
by their tension, and thus by $g=4\pi$  and $q$,
\beq
\cos(\pi-\omega)= \frac{g}{\sqrt{g^2+q^2}}.
\eeq
This angle $\omega$ is depicted in Fig. 1.

\vskip 5mm
\begin{center}
\leavevmode
\epsfysize =1.5in\epsfbox{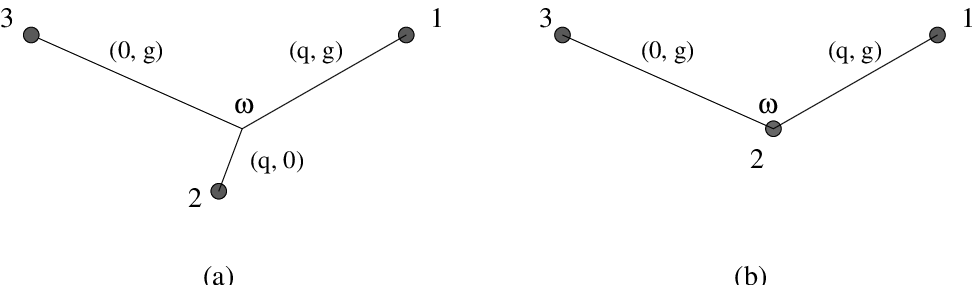}
\end{center}
\vskip 0mm
\begin{quote}
{\bf FIG. 1.} Configurations of Three-Pronged Strings when it is
(a) stable or (a) at the threshold of instability. We labeled the 
D3-branes by numeral 1,2,3 in accordance with the choice of basis
in section 3.
\end{quote}

The three-pronged string becomes marginally (un)stable whenever
any one of the string has zero length. This happens either due to change
of Higgs vev's or due to  change in electric charge/coupling.
In Fig. 1,  we described the case where the Higgs vev's change.
When the fundamental string become arbitrarily short so that
the second D3-brane coincides with the junction at the
center, the string configuration is made only of $(0,g)$ and $(q,g)$ 
strings. The Higgs force is still attractive but not strong enough compare
with the repulsive force from the presence of the relative electric
charge; the system is no longer classically bound. In this limit, the angle
between $X^I_{21}$ and $X^I_{32}$ must become $\pi-\omega$. Indeed 
it is not hard to show that
\beq
\frac{{X}_{21} \cdot {X}_{32}}{|{X}_{21}||{X}_{32}|}
\le \cos(\pi-\omega),
\eeq
where the equality holds precisely when Higgs vev's and electric
charge are such that $R\rightarrow \infty$.
Thus we find the same instability in both string and field theory pictures.

There are other kind of instability, for instance, when $(q,g)$ string
becomes arbitrarily short. Clearly there is no static electromagnetic 
force between the electric and magnetic charges. In this case, the
cause of instability in field theoretical term, turned out be due to
the repulsion from the Higgs interaction. This is the limit where
$\mu_1=h_2-h_1=0$ in the field theory, and where $X^I_{12}$ and
$X^I_{13}=X^I_{12}+ X^I_{23}$ becomes mutually orthogonal in the
string picture.

\section{$1/4$-BPS Dyons from Quantum Excitations}

In principle, the supermultiplet structure of the $1/4$-BPS states  should
be recovered from low energy quantum mechanics of the above solitonic
solution. However, in this paper, we will take a shortcut, and ask the
question of degeneracy by presenting an alternate construction of these
dyonic states. For simplicity, we will confine the present discussion to 
the case of $SU(3)$.

We start with the spherically symmetric magnetic monopole solution obtained
by an $SU(2)$ embedding along the root $\balpha+\bbeta$ with the single
nonuniform Higgs $b\cdot\phi$.  If $a\cdot\phi$ vanished, the monopole would 
have 8 bosonic and 8 fermionic zero modes. In a generic vacuum where $\langle
a\cdot\phi\rangle \neq 0$, however, half of these 16 zero modes are lifted
and acquire finite energy. Of the remaining 4 bosonic zero modes, three
corresponds to translations and one is generated by global $U(1)$ 
transformations. There are also 4 fermionic zero modes, quantization of 
which imparts a $N=4$ vector multiplet structure, thus the
degeneracy $2^4$, to the soliton.

A minimal $1/4$-BPS states should have a degeneracy factor of $2^6$
and highest spin $3/2$. To see how such structures arise, we need to pay
close attention to those modes lifted by $\langle a\cdot\phi\rangle
\neq 0$. Fermionic modes are easiest to follow.  Introduce a basis for
Dirac matrices where $\gamma^0$ is diagonal and $\gamma^5$ is
off-diagonal,
\begin{eqnarray}
\gamma^0&=&-i\otimes \sigma^3, \\
\gamma^k&=&\sigma^k \otimes \sigma^2, \\
\gamma^5&=&1\otimes \sigma^1,
\end{eqnarray}
with 2 by 2 Pauli matrices $\sigma^i$'s. Using $SO(6)$ R-symmetry, one
can bring the Dirac equation to the following form,
\beq
\gamma^0\left[i\gamma^k D_k+ \gamma^5 \, b\cdot\phi \pm i\,
a\cdot\phi\right]\Psi_\pm= \epsilon\,\Psi_\pm,
\label{dirac}
\eeq
written in the time-independent form with the energy eigenvalue $\epsilon$. 
Here we used a static gauge with the purely magnetic background solution.
$N=4$ theory has two (adjoint) Dirac fermions, which together lift to a Dirac
spinor in 6-dimensions. The two are of opposite six-dimensional chiralities,
and the subscript $\pm$ refers to this fact.

Decomposing the Dirac spinors as $\Psi =(\chi,\psi)^T$ in term of
two-component spinors, and defining an operator ${\cal D}\equiv
i\sigma^kD_k+i\,b\cdot\phi$, the Dirac equations is rewritten as,
\beqn
{\cal D}\psi_\pm \pm \,[a\cdot\phi,\chi_\pm] &=& \epsilon\chi_\pm ,\\
{\cal D}^\dagger\chi_\pm \mp \, [b\cdot\phi,\psi_\pm] &=& \epsilon\psi_\pm .
\eeqn
Recall that, given a BPS background monopole configuration that
satisfies $B_k=D_k (b\cdot\phi)$, the operator ${\cal D}$ has zero
modes while ${\cal D}^\dagger$ does not.  When $a\cdot\phi =0 $, each
Dirac fermion contributes 4 zero modes ($E=0$); they solve ${\cal
D}\psi=0$ and $\chi =0$. The 4 solutions to ${\cal D}\psi=0$ can be
labeled by the representation under the embedded $SU(2)$. The adjoint
representation of the gauge group $SU(3)$ is decomposed into a
triplet, a pair of doublet, and a singlet with respect to the $SU(2)$
embedded along $\balpha+\bbeta$. The singlet is associated with the
generator $\balpha\cdot{\bf H}-\bbeta\cdot{\bf H}$, while the two
doublets are associated with the pairs $(E_\alpha,E_{-\beta})$ and
$(E_\beta,E_{-\alpha})$. The triplet would contribute two zero modes,
and each doublet would contribute one, which account for all four
solutions to ${\cal D}\psi=0$.

By construction of Eq.~(\ref{dirac}), the uniform field $a\cdot\phi$
is orthogonal to the total magnetic charge $\balpha+\bbeta$;
\beq
a\cdot\phi = v\,(\balpha\cdot{\bf H}-\bbeta\cdot{\bf H}),
\eeq
which has a nontrivial commutator only with isospin doublets, and even
then acts on each as an multiplication by a number.  With
$a\cdot\phi\neq 0$, therefore, those modes from the isospin triplets
commutes with $b\cdot \phi$ and survives as zero modes. As mentioned
above, quantization of these leads to a vector multiplet structure of
degeneracy $2^4=16$.

The other four from isospinor doublets can no longer be zero modes, 
however, and are promoted to finite energy eigenmodes of the form
\cite{mans},
\beq
\Psi_\pm = e^{-i\epsilon t}\left(\matrix{0 \cr \psi}\right).
\eeq
The isospin doublet, 2-component spinor $\psi$ is exactly of the same mode 
that solves ${\cal D}\psi=0$, and thus are normalizable. They are compactly
supported around the monopole core. The energy eigenvalue $\epsilon$ equals 
$\pm 3v/2$ for the first  doublet and $\mp 3v/2$ for the second 
doublet. This is because
\begin{equation}
[a\cdot\phi, E_\alpha]=\frac{3v}{2} E_\alpha,\qquad
[a\cdot\phi, E_{-\beta}]=\frac{3v}{2} E_{-\beta},
\end{equation}
and similarly for $E_\beta$ and $E_{-\alpha}$ with a negative sign. 
Filling the Dirac sea up to $\epsilon=0$, creation (or annihilation) of one of 
these eigenmodes will result in a 
quantum excitation that costs a positive energy $|\epsilon|=|3v/2|$.

To check against the BPS mass formula, we need the behavior of electric field
at large distances when one of these modes is turned on. From various
considerations, it is well known that these modes from gauge doublets carry no 
angular momentum. This can be surmised from the angular momentum formula,
${\bf J}={\bf L}+{\bf s}+{\bf t}$, where the $SU(2)$ gauge generators $\bf t$
are added to orbital and spin angular momenta. The solution to ${\cal D}
\psi=0$ with an $SU(2)$ doublet $\psi$ is unique and spherically symmetric 
(${\bf L}^2=0$), hence must be of the form,
\beq
\psi_\pm \quad\propto\quad \frac{1}{\sqrt{2}}|E_\alpha, s_z=-1/2\rangle 
-\frac{1}{\sqrt{2}}|E_{-\beta}, s_z=+1/2\rangle ,
\eeq
from the first doublet, and
\begin{equation}
\psi_\pm \quad\propto \quad\frac{1}{\sqrt{2}}|E_\beta, s_z=-1/2\rangle 
-\frac{1}{\sqrt{2}}|E_{-\alpha}, s_z=+1/2\rangle  ,
\end{equation}
from the second. The isospin and the spin are correlated in such a way
that ${\bf J}^2=({\bf s}+{\bf t})^2=0$. From this, we learn that the
mode by itself carries an electric charge of $ \pm
(\balpha-\bbeta)/2$, or the relative charge is $\Delta q = \mp 1/2$.

However, there is a well known subtlety associated with turning on
such a mode from a gauge doublet. Because it acquires a phase of $-1$ upon a
gauge rotation corresponding to the center of $SU(2)$, its excitation
must be accompanied by a half-integer momentum along internal phase
angle of the background monopole. This leads to additional electric
charges of the form $(m/2)(\balpha+\bbeta)$ for any odd integer
$m$. The minimal states are those with $m=\pm 1$. Combining this with
the fermionic contribution, we find the electric charges are $\pm
\balpha$ or $\mp\bbeta$. With two Dirac spinors $\Psi_\pm$ then,
quantization leads to eight minimal dyonic excitations, which split
into four pairs of identical electric charges, $\balpha$, $-\bbeta$,
$\bbeta$, $-\balpha$.  Excitation energy due to the half-integer
momentum $m/2=\pm 1/2$ is of second order in electric charge, and will
not affect the leading approximation.

Does the leading excitation energy $|\epsilon|=|3v/2|$ agree with 
the general BPS mass formula? In the limit of small electric 
coupling,\footnote{We remind readers that $Q^E$ has a factor of $e$ while
$Q^M$ has a factor of $1/e$. We suppressed $e$ from notations
in section 3 and thereafter.} the central charges may be expanded as
\begin{equation}
Z_\pm=\sqrt{(Q^M)^2+(Q^E)^2\pm 2Q^M Q^E \sin\alpha}\simeq
Q^M\pm Q^E\sin\alpha+\cdots.
\end{equation}
The actual BPS bound is Max($Z_+,Z_-$), so the first order correction 
due to the electric charge is
\begin{equation}
| Q^E\sin\alpha | \simeq|{\rm tr}\left((a\cdot\phi) ({\balpha}\cdot {\bf H})
\right)|=|{\rm tr}\left((a\cdot\phi) ({\bbeta}\cdot {\bf H})
\right)|=\left|\frac{3v}{2}\right| ,
\end{equation}
which coincides with $|\epsilon|=|3v/2|$, as it should if the dyonic
state is indeed $1/4$-BPS. The bosonic counterpart of this eigenmode
analysis should proceed similarly, except that the corresponding
eigenmodes will come in a pair of spin doublets rather than four spin
singlets. The final result is, then, for each electric charge,
$\balpha$, $-\bbeta$, $\bbeta$, $-\balpha$, there are $2+2=4$ dyonic
excitations due to the gauge-doublet eigenmodes: the net degeneracy of
the resulting dyon is $4\times 2^4= 2^6$ for each electric charge,
where we take into account the extra degeneracy of $2^4$ due to the
four fermionic zero modes from $SU(2)$ triplets. The spin content of
each dyon multiplet is that of two $N=4$ vector multiplets (from
fermionic eigenmodes) plus a tensor product of a spin doublet and one
$N=4$ vector multiplet (from bosonic eigenmodes). This is precisely
the $1/4$-BPS multiplet of highest spin $3/2$. The four types of
$1/4$-BPS dyons correspond to the four different string configurations
depicted in Fig. 2. 

\vskip 1cm
\begin{center}
\leavevmode
\epsfysize=3.5in
\epsfbox{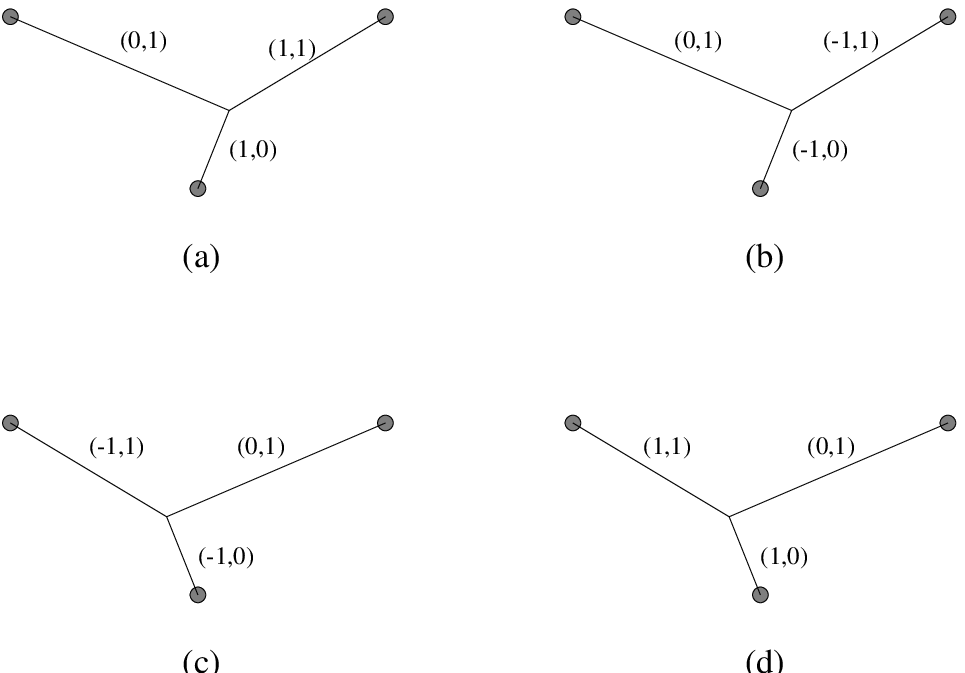}
\end{center}

\begin{quote}
{\bf FIG. 2.} Four different minimal dyonic states of magnetic charge
$\balpha+\bbeta$. Electric charges are respectively (a) $\balpha$, (b) 
$-\balpha$, (c) $\bbeta$, and (d) $-\bbeta$. For a match with standard
notations in string theory, in this figure we  relabeled the unit
D-string by $(0,1)$, instead of $(0,4\pi)$.
\end{quote}

Some discussion is due on the validity of the approximation. Note that
the expansion of the BPS mass formula proceeded with the assumption,
\begin{equation}
Q^M \gg Q^E\sin\alpha \gg \frac{(Q^E\cos\alpha)^2}{Q^M},
\end{equation}
which is obtained by expanding the BPS bound. It is clear from the 
subleading contributions to dyon energies that these criteria are necessary 
for a successful match between the BPS mass and the energy found from 
eigenmode analysis. The first condition simply says that the 
excitation energy should be much smaller that the mass of the bare soliton 
itself, and is to be expected. What does the second condition do?

The present approximation takes into account only part of the backreactions.
It does address the change in long-range electric field in response to the
excitation, but ignored its counterpart in magnetic soliton structure.
This is of course why we seem to obtain spherically symmetric configuration,
even though we clearly demonstrated that this should happen rarely in
exact dyonic states. The consequence is that our choice of $b_I$ is 
independent of the electric charge being turned on, such that $b_I$ is in
fact parallel to $Q_I^M$. To obtain the correct BPS bound, in reality,
the angle $\theta$ between $b_I$ and $Q_I^M$ must be given by
\begin{equation}
\tan\theta=\frac{\pm Q^E\cos\alpha}{Q^M\pm Q^E\sin\alpha}\simeq
\pm\frac{Q^E\cos\alpha}{Q_M},
\end{equation}
where we used the first condition $Q^M\gg Q^E\sin\alpha$. The BPS bound,
\begin{equation}
b_IQ^M_I+a_IQ^E_I,
\end{equation}
then contains an error of order
\begin{equation}
\delta\theta^2Q^M\pm\delta 
\theta Q^E\cos\alpha \sim \frac{(Q^E\cos\alpha)^2}{Q^M},
\end{equation}
where $\delta\theta'\equiv\theta-\theta'=\theta$, due to the incorrect
angle $\theta'=0$. Since we ignore the magnetic backreaction to the
quantum excitation, we must require this error be negligible against
the first order estimate, which explains the second condition.  It
also explains why we do not find the phenomenon of instability in the
present set-up. Bergman's criteria tells us that it occurs when $(Q^E
\cos\alpha)^2$ is comparable to $Q^M Q^E\sin\alpha$, where magnetic
backreaction to the quantum excitations are of a first order effect,
instead of being a second order effect. Instability cannot be probed
without taking into account the reaction of magnetic soliton to the
quantum excitation. In this sense, the two constructions we gave are
complimentary to each other; the first gave us the understanding of
the dynamics while the second is better suited for state counting.

\section{Degeneracy and Supermultiplet Structure of Dyons}

In the previous section, we saw how the supermultiplet of degeneracy $2^6$
arises in case of minimally charged BPS states. The method we developed is 
applicable for 1/4-BPS states  with higher electric charges, and we
will summarize 
the general supermultiplet structure. Let us parameterize the quantized
electric charge by writing
\beq
{\bf q}=q_\alpha\balpha+q_\beta\bbeta=
\frac{k}{2}\,(\bbeta-\balpha)+\frac{m}{2}\,(\balpha+\bbeta).
\eeq
with integers $k$ and $m$. Consistent quantization requires that $m$
is odd(even) whenever $k$ is odd(even).  The relative charge of the
system is given by $\Delta q=(q_\beta-q_\alpha)/2=k/2$. The integer
$k$ corresponds to the number of excited eigenmodes while $m/2$ is the
momentum along a internal $U(1)$ angle of the magnetic solitons. The
case of no relative electric charge $\Delta q=0$ corresponds to the
usual BPS dyon that breaks half of supersymmetry, which come in an
$N=4$ vector multiplet. The case of $\Delta q=\pm 1/2$ was addressed
in the previous section. The supermultiplet structure found there can
be summarized in terms of the eigenvalues under one of the angular
momentum operators, $J_3$,
\vskip 5mm
\begin{center}
\begin{tabular}{|c|c|c|c|c|c|c|c|}
\hline
$J_3$ & 3/2 & 1 & 1/2 & 0 & -1/2 & -1 & -3/2 \\
\hline
Degeneracy & 1 & 6 & 15 & 20 & 15 & 6 & 1\\
\hline
\end{tabular}
\end{center}
The total degeneracy is $2^6$, which, for $1/4$-BPS state, is the smallest 
while being also consistent with supersymmetry. Call this multiplet $G_0$.
This multiplet can be seen as a tensor product between $N=4$ vector 
multiplet with a $N=1$ chiral multiplet.

Higher charged states with $|\Delta q|\ge 1$ is obtained by exciting
appropriate eigenmodes $k=2|\Delta q|$ times. Given a fixed electric
charge, there are always two bosonic and two fermionic eigenmodes at
disposal. There are $k+1$ states where no fermionic modes
are excited, $2k$ states where one fermionic modes are
excited, and $k-1$ states where both fermionic modes are
excited.  Combining the degeneracy from four fermion zero modes of the
center of mass motion, we then find the total degeneracy of $4k\times 2^4=4
(2|\Delta q|)\times 2^4=(2|\Delta q|)\times 2^6$. For detailed spin
content, we only need to recall that $2^4$ has the vector structure
and that bosonic excitations carry extra spin of $\pm 1/2$. The result
is the sum of $2|\Delta q|$ tables identical to the above, except that
$J_3$ eigenvalues are shifted,
\vskip 5mm
\begin{center}
\begin{tabular}{|c|c|c|c|c|c|c|c|}
\hline
$J_3- S $ & 3/2 & 1 & 1/2 & 0 & -1/2 & -1 & -3/2 \\
\hline
Degeneracy & 1 & 6 & 15 & 20 & 15 & 6 & 1\\
\hline
\end{tabular}
\end{center}
with $S$ ranging from $-|\Delta q|+1/2$ to $|\Delta q|-1/2$ in step of 1.  
The resulting supermultiplet has a tensor product
structure $G_0\otimes [|\Delta q |-1/2]$ where we denoted by $[|\Delta
q|-1/2]$ the spin $|\Delta q| -1/2$ representation of the angular momentum. 
The highest spin of such a multiplet is $|\Delta q|+1$. From construction,
it is easy to see that $|\Delta q|$ of this arises from bosonic excitations.
The only fermionic contribution comes from the four fermionic zero modes,
which tops out at $1$. 

This bosonic spin has a rather interesting explanation in the context
of classical dyonic configurations of section 3. Consider the limit of
large Higgs vev's. In this limit, the solution degenerates to a pair of
point-like dyons of $\balpha$ and $\bbeta$ type, each carrying electric 
charges $q_\alpha$ and $q_\beta$.  The conserved angular 
momentum is known to contain an anomalous contribution in this situation,
\beq
{\bf J}={\bf L}+\frac{g\Delta q}{4\pi}\hat{\bf R}
\label{point}
\eeq 
proportional to the relative electric charge $\Delta
q=(q_\beta-q_\alpha)/2$~\cite{manton}.  The unit vector $\hat{\bf R}$
points from $\balpha$ dyon to $\bbeta$ dyon. With the unit magnetic
charges, $g=4\pi$, the anomalous angular momentum is exactly $|\Delta
q|$, as expected. (We fully expect that a classical field theoretic
calculation of the anomalous angular momentum for the 1/4-BPS
configurations will reproduce the answer (\ref{point}) obtained in the
point-like dyon limit. See Appendix E for a simple expression for the
angular momentum.)

\section{Conclusion}

In this paper we explored 1/4-BPS states in $N=4$ supersymmetric
theories which correspond to three-pronged strings ending on D3-branes
in Type IIB string theory. 1/4-BPS configurations typically consist of
two (or more) dyonic cores, which are positioned so that static
electromagnetic  force is perfectly balanced against scalar
Higgs force. The marginal instability previously found in string
picture is shown to arise from the excessive repulsion from either
electromagnetic or Higgs interaction. An alternate construction using the
finite energy excitations around purely magnetic soliton also revealed
supermultiplet structure of 1/4-BPS states with arbitrary relative
electric charge.  The degneracy and the highest spin the
supermultiplets grow linearly with the relative charge. In the minimal
cases, the multiplet has the degeneracy of $2^6$ with the highest spin
$3/2$.

In principle, the question of degeneracy and supermultiplet structures
can also be addressed by considering low energy quantum mechanics of
the classical $1/4$-BPS solution we found. This would necessarily involve
zero-mode analysis of these nonspherical solitons, which we did not
attempt.

Our constructions can be  generalized to the case of multi-pronged 
string configurations in larger gauge groups. In the small coupling limit,
the same eigenmode analysis should produce the dyonic states of higher
magnetic and electric charges. Also classically, one can distribute many 
monopole in the background, and solve for possible electric configurations.
We expect to find multi-dyon configurations hung together by
the delicate balance of static forces. We should be able to exploit
the ADHMN formalism as in this work to explore these field configurations. 
One interesting case is when the gauge symmetry is partially restored as 
in Ref.~\cite{nonabelian}. For solutions whose net magnetic charge 
is  abelian, the configuration typically consists of massive magnetic 
cores surrounded by nonabelian magnetic clouds. It would be interesting
to see if any new physics arises by considering $1/4$-BPS version of
such nonabelian configurations.

While we considered only $N=4$ theories so far, it is
clear that the methods developed here can be applied to $N=2$
theories with minimal modifications. $N=2$ supersymmetry 
algebra possesses half the supersymmetry generators and also 
only one central charge, so we naturally expect the 
spectrum be qualitatively different. This is quite apparent
from the point of view adopted in section 5, since reducing 
supersymmetry involves removing one of the two adjoint Dirac 
spinors. In fact, there appears to be no guarantee that the
present constructions produce proper $1/2$-BPS states. It may in
general depend on the particular electromagnetic charges, 
Higgs vev's, and other details of the theory. We
are currently exploring some of the issues.

\vskip 5mm

As this work was being completed, two related papers 
\cite{multi,kol} have appeared.

\vskip 1cm

\centerline{\bf Acknowledgments} 

We thank Philip Argyres, Cumrun Vafa, Thomas Kraan, Hugh Osborn and
Erick Weinberg for useful discussions. K.L. is supported in part by
the U.S. Department of Energy. P.Y. is supported in part by the
National Science Foundation.

\newpage

\vskip 1cm
\setcounter{equation}{0}
\makeatletter
\renewcommand\theequation{A\arabic{equation}}
\makeatother

\leftline{\large\bf Appendix A: The ADHM Formalism}
\vskip 3mm
\noindent
The ADHM formalism~\cite{adhm} for $k$ instantons of the SU(n) gauge theory
starts with a $(n+2k)\times 2k$ matrix
\beq
\Delta = \left(\matrix{\lambda_{n\times 2k} \cr \mu_{2k\times 2k}
\cr}\right) + \left(\matrix{0\cr I_{2k\times 2k} \cr}\right)x,
\eeq
where $x=x_\alpha e_\alpha$ and $e_\alpha=(i\sigma_j, 1)$~\cite{adhm}.
Finding the $(n+2k) \times 2k$ matrix $v$ such that
\beq
\Delta^\dagger v = 0,\;\;\;\; v^\dagger v = I_{n\times n},
\label{adhm1}
\eeq
we can construct the anti-hermitian gauge field 
\beq
A_\alpha = v^\dagger \partial_\alpha v.
\eeq
The condition for the field strength to be self-dual is
that 
\beq
(\Delta^\dagger \Delta)_{2k\times 2k} = f^{-1}_{k\times k } I_{2\times
2}.
\eeq
This implies that $\mu=\mu_\alpha e_\alpha$ with  hermitian matrices
$(\mu_\alpha)_{k\times k}$ and that
\beq
i\eta_{\alpha\beta}^i [\mu_\alpha, \mu_\beta] + tr_2 ( \sigma^i
\lambda^\dagger \lambda) =0,
\label{nahm0}
\eeq
where $e^\dagger_\alpha e_\beta = \delta_{\alpha\beta} +
i\eta_{\alpha\beta}^i\sigma^i$ with anti-self-dual `t Hooft tensor
$\eta^i_{\alpha\beta}$. 
The inverse $k\times k$ matrix  $f$ satisfies equation
\beq
\biggl\{ (\mu_\alpha +x_\alpha)^2 +\frac{1}{2} \tr_2 \lambda^\dagger
\lambda \biggr\} f = I_{k\times k}.
\eeq

We can choose the $v$ such that
\beq
v_{(n+2k)\times n} = \left(\matrix{I_{n\times n} \cr u_{2k\times n}
\cr} \right) N^{-\frac{1}{2}},
\eeq
where $N= 1+u^\dagger u $ is an $n\times n$ hermitian matrix~\cite{kraan}.
The ADHM equation becomes 
\beq
 (\mu^\dagger+x^\dagger) u +\lambda^\dagger = 0.
\label{adhm2}
\eeq
The gauge field becomes
\beq
A_\alpha = N^{-\frac{1}{2}} (u^\dagger \partial_\alpha u )N^{-\frac{1}{2}}+
N^{-\frac{1}{2}}\partial_\alpha N^{-\frac{1}{2}}.
\eeq
The self-dual field strength is then given by
\beq
F_{\alpha \beta} = 2i  N^{-\frac{1}{2}}u^\dagger f \bar{\eta}_{\alpha\beta} u
N^{-\frac{1}{2}}, 
\eeq
where $e_\alpha e^\dagger_\beta= \delta_{\alpha\beta} + i
\bar{\eta}_{\alpha\beta} $ where $\bar{\eta}_{\alpha\beta}$ is the self-dual
't Hooft tensor.

The construction has redundancy,
\beq
 \lambda \rightarrow  \lambda U,\;\;\; \mu\rightarrow U^\dagger \mu
U, \;\;\; u \rightarrow  U^\dagger u,
\eeq
where $U$ belongs to $U(k)$. The number of parameters of $\mu_\alpha$ and
$\lambda$ are
\beq
\mu_\alpha: 4k^2,\;\;\;\; \lambda:4nk.
\eeq
The number of the  conditions (\ref{nahm0}) are  $3k^2$ and the number
of $U(k)$ elements is $k^2$. Thus the net number of independent
variables for a $k$ instanton in $SU(n)$ is
\beq
4k^2+4nk - 3k^3-k^3 =4nk.
\eeq

\vskip 1cm
\setcounter{equation}{0}
\makeatletter
\renewcommand\theequation{B\arabic{equation}}
\makeatother

\leftline{\large\bf Appendix B: The Nahm  Formalism of Calorons}
\vskip 3mm
\noindent

We consider instanton solutions on $R^3\times S^1$ with nontrivial
Wilson loop, which can be regarded as the infinite infinite number of
instantons which is quasi-periodic along $x_4$
axis~\cite{caloron,piljin,kraan}. We analyze these calorons by
extending the method  in Ref.~\cite{kraan} to the case of  $SU(n)$
gauge group, along the way, by connecting to the Nahm's
formalism~\cite{nahm}.  We choose the unit interval of the $x_4$ to be
$[0,\beta]$ and imagine the number of instantons in a given interval
is $k$. The ADHM matrices becomes
\beq
\Delta(x) = \left(\matrix{\lambda_l \cr \mu_{l l'} \cr}\right) + 
\left(\matrix{0\cr x\delta_{l l'}\cr}\right),
\eeq
where $l,l'$ are integers. Here $\mu_{ll'}$ for each $ll'$ is 
a $2k\times 2k$ matrix and $\lambda_l$ for each $l$ is a $2k\times n$
matrix. 

We consider  the gauge field to be quasi-periodic so that 
\beq
A_\alpha({\bf x},x_4+\beta)= e^{i\beta {\bf h}\cdot
{\bf H}} A_\alpha({\bf x},x_4) e^{-i\beta {\bf h}\cdot {\bf H}}. 
\eeq
This is equivalent to considering the periodic field configurations
with the asymptotic value at spatial infinity to be
\beq
<A_\alpha> = i{\bf h}\cdot {\bf H} \delta_{\alpha 4}.
\eeq
Note that ${\bf h}\cdot {\bf H} = \sum_{a=1}^n h_a P_a$ such that
$\sum_a h_a = 0 $ with  $P_a$ being the projection operator to the $a$
component of any $n$-dimensional vector. We can choose the gauge so
that 
\beq
h_1<h_2<...<h_n<h_1+\frac{2\pi}{\beta}.
\eeq

The condition (B2) can be satisfied if
\beq
u_l({\bf x},x_4+\beta) =  u_{l-1}({\bf x},x_4)   e^{-i\beta {\bf
h}\cdot {\bf H}}  ,
\eeq
which in turn can be satisfied if 
\beqn
& & \lambda_l^\dagger = 
\lambda^\dagger_{l-1}   e^{-i\beta {\bf h}\cdot {\bf H}}, \\ 
& & \mu_{ll'}= \mu_{(l-1)(l'-1)} - \beta e_4 \delta_{ll'}.
\eeqn
These relations lead to 
\beqn & & \lambda_l^\dagger =\lambda_0^\dagger  e^{-i\beta l{\bf
h}\cdot {\bf H}} , 
 \\ &
& \mu_{ll'}^\alpha= T_{ll'}^\alpha - l \beta \delta_{\alpha 4},
\delta_{ll'},
\eeqn
such that $T_{ll'}^\alpha = T_{(l-1)(l'-1)}^\alpha$.
Note that $(\Delta^\dagger \Delta )_{ll'}(x_4+\beta) = (\Delta^\dagger
\Delta )_{(l-1)(l'-1)}(x_4)$ and so $f_{ll'}(x_4+\beta) =
f_{(l-1)(l'-1)}(x_4)$.

We introduce the Fourier transformation of these matrices:
\beqn
& & \lambda^\dagger(t) = \sum_l e^{i\beta tl} \lambda^\dagger_l, \\ 
& & T_\alpha (t) = \sum_l e^{i\beta t l } T^\alpha_{l0} ,\\
& & u(t) = \sqrt{\frac{\beta}{2\pi}}\sum_l e^{i\beta tl} u_l, \\
& & f(t,t') = \frac{\beta}{2\pi}\sum_{ll'} e^{i\beta tl} f_{ll'}
e^{-i\beta t'l'} .
\eeqn
Note that $T_\alpha(t)$ is hermitian $k\times k$ matrix and periodic
under $t\rightarrow t+2\pi/\beta$, $\lambda^\dagger(t)$ is $n\times
2k$ and periodic, and $u(t)$ is $n\times 2k$ and periodic. The
function $f(t,t')$ is  periodic under
shift of $t, t'$ with $2\pi/\beta$.

Furthermore, from Eqs.~(B8) and (B10), we get
\beq
\lambda^\dagger(t) = \frac{2\pi}{\beta} \lambda^\dagger_0 \sum_a
\delta(t-h_a) P_a .
\eeq
{}From the property that $u(t,x_4+\beta)= u(t,x_4)e^{i\beta(t-{\bf
h}\cdot{\bf H})},$ we can introduce 
\beq
u_* (t;{\bf x},x_4) =
u(t;{\bf x},x_4 )e^{-ix_4(t-{\bf h}\cdot {\bf H})},
\eeq
such that $u_*(t+2\pi/\beta) = u_*(t)e^{i 2\pi x_4/\beta}$ and
$u_*(x_4+\beta)= u_*(x_4)$.

In the Fouriered functions, the consistent condition (\ref{nahm0})
becomes the Nahm equation for a caloron~\cite{murray,caloron},
\beq
\partial_t T_i - i[T_4,T_i] = \frac{i}{2} \epsilon_{ijk}[T_j,T_k] + 
\frac{1}{2} \tr_2 \sigma_i w^\dagger\sum_a\delta(t-h_a)P_a w,
\eeq
where $w=\sqrt{2\pi/\beta}\lambda_0$. 
The ADHMN equation (\ref{adhm2}) for $u(t)$ becomes 
\beq
\biggl[e_4^\dagger(i\partial_t +T_4 +x_4) +e_i^\dagger
(T_i+x_i)\biggr] u(t) + 
w^\dagger \sum_a \delta(t-h_a)P_a =0.
\label{adhm3}
\eeq
In terms of the quasi-periodic  $u_*(t)$, the above equation becomes
\beq
\biggl[i\partial_t+ T_4  -i\sigma_i (T_i+x_i)\biggr] u_*(t) +
w^\dagger \sum_a \delta(t-h_a)P_a =0.
\label{adhm4}
\eeq
This is the standard Nahm equation for magnetic monopoles~\cite{nahm}.

In this process  the normalization factor $N^{-\frac{1}{2}}$ becomes 
\beq
N^{-\frac{1}{2}} =  e^{i{\bf h}\cdot {\bf H} x_4} N_*^{-\frac{1}{2}}
 e^{-i{\bf h}\cdot {\bf H} x_4},
\label{nstar}
\eeq
where $N_* = 1+ \int_0^{2\pi/\beta}dt u_*^\dagger u_* $ is
single-valued under $x_4\rightarrow x_4+\beta$.  After singular gauge
transformation $e^{i{\bf h}\cdot {\bf H}x_4}$, the gauge field becomes
single-valued and is given by
\beqn
& & A_{*4} = N_*^{-\frac{1}{2}} i\sum_a h_a P_a N_*^{-\frac{1}{2}}
+ N_*^{-\frac{1}{2}} \int_0^{2\pi/\beta} dt \, it u_*^\dagger(t) u_*(t)
N_*^{-\frac{1}{2}} , \nonumber \\
& & A_{*i} =  N_*^{-\frac{1}{2}} \partial_i  N_*^{-\frac{1}{2}}
+ N_*^{-\frac{1}{2}} \int_0^{2\pi/\beta} dt u_*^\dagger(t) 
\partial_i  (u_*(t) N_*^{-\frac{1}{2}}),
\eeqn
which is the standard form of the Nahm construction for the self-dual
magnetic monopoles~\cite{nahm}.

We redefine the Green function  $f_*(t,t;x_4) = e^{-ix_4
t}f(t,t';x_4)e^{ix_4 t'}$, which is single-valued in $x_4$ but 
multi-valued in $t$. It satisfies
\beq
(i\partial_t +T_4)^2 f_* + (T_i +x_i)^2 f_* + 
\frac{1}{2} W(t) f_* = \delta(t-t'),
\eeq
where
\beq
W(t)=\tr w^\dagger \Sigma_a \delta(t-h_a)P_a w.
\eeq
The single-valued self-dual field strength becomes
\beq
F_{*\alpha\beta}=   N_*^{-\frac{1}{2}}\left\{  \int dt dt'
u_*^\dagger(t)f_*(t,t') 
\bar{\eta}_{\alpha\beta} u_*(t') \right\} N_*^{-\frac{1}{2}}.
\eeq

\vskip 1cm
\setcounter{equation}{0}
\makeatletter
\renewcommand\theequation{C\arabic{equation}}
\makeatother

\leftline{\large\bf Appendix C: The Adjoint Scalar Field}
\vskip 3mm
\noindent
The general method to find  the solution of the covariant Laplacian
for a scalar field in the adjoint representation has been developed in
the instanton background~\cite{osborn2}. We start with a general form
\beq
\Phi(x) = v^\dagger Q v,
\eeq
where $Q$ is an hermitian  $(n+2k)\times(n+2k)$ matrix. 
We assume that $Q$ is independent of $x$ and takes the ansatz
\beq
Q=\left(\matrix{ q_{n\times n} & 0 \cr 0 & p_{k\times k} I_{2\times 2}
\cr}\right) .
\eeq

Using the fact that the projection operator $P=vv^\dagger= I-\Delta f
\Delta^\dagger$, one can show that 
\beq
D_\alpha^2 \Phi = 4N^{-\frac{1}{2}}
u^\dagger f \biggl[ \tr_2\biggl(\lambda^\dagger q
\lambda-\frac{1}{2}\{\lambda^\dagger\lambda,p\}\biggr) -
[\mu_\alpha,[\mu_\alpha,p]] \biggr] fu N^{-\frac{1}{2}},
\eeq
where $\tr_2$ is a trace over 2-dimensional part of matrices.
With two hermitian $k\times k$ matrices,
\beq
W= \tr_2 \lambda^\dagger \lambda, \;\;\; \Lambda =
\tr_2\lambda^\dagger q \lambda,
\eeq
the condition for the scalar field to satisfy the covariant Laplace
equation $D_\alpha^2 \Phi =0$ becomes a condition on the matrix $p$, 
\beq
-[\mu_\alpha,[\mu_\alpha,p]]-\frac{1}{2}\{W,p\}+\Lambda = 0 .
\label{peq1}
\eeq
Note that the above equation determines $p$ for a given infinitesimal
generator $q$ of $SU(n)$.
Especially when $q=I_{n\times n}$, we can see $p=I_{k\times k}$ solves
the above equation.

For similar scalar fields in any caloron background, we extend the method
described in Appendix B. We generalize Eq.~(C2) to an infinite
dimensional matrix, and then the analogy of Eq.~(C1) would be
\beq
\Phi = N^{-\frac{1}{2}}q N^{-\frac{1}{2}} +  N^{-\frac{1}{2}}
u^\dagger_l p_{ll'} u_{l'}  N^{-\frac{1}{2}}. 
\eeq
Similar to  the gauge field,  the adjoint Higgs  scalar field should
satisfy the
quasi-periodic condition $\Phi({\bf x}, x_4+\beta) = e^{i\beta{\bf
h}\cdot {\bf H}}\Phi({\bf x},x_4)e^{-i\beta{\bf h}\cdot {\bf H}}$.
Thus the above ansatz is consistent with Eq.~(\ref{nstar}) only if 
\beq
[{\bf h}\cdot {\bf H}, q]=0.
\eeq
This equation implies that 
there are only $n-1$ independent $q$'s when the gauge symmetry is
maximally broken or all $h_a$ are different.

To consider the similar solution around magnetic monopoles, we again
Fourier transform $p$ matrix, 
\beq
p(t) = \sum_l e^{i\beta tl } p_{l0}.
\eeq
Then, we can re-express Eq.~(\ref{peq1}) as an ordinary differential equation
for $k\times k$ hermitian matrix $p(t)$,
\beq
[\partial_t -iT_4,[\partial_t-iT_4,p(t)]]-[T_i(t), [T_i(t),p(t)]] -
\frac{1}{2} \{ W(t),p(t)\}
+ \Lambda(t) = 0,
\label{pstar}
\eeq
where $W(t) = \tr_2 w^\dagger \sum_a \delta(t-h_a) P_a w$ and
$\Lambda(t) = \tr_2 w^\dagger \sum_a \delta(t-h_a) P_a q w$.  For such
a solution $p(t)$, after a gauge transformation by $e^{-ix_4{\bf
h}\cdot {\bf H}}$, the single-valued solution of adjoint scalar
Laplace equation is given by
\beq
\Phi_* = N_*^{-\frac{1}{2}} q N_*^{-\frac{1}{2}} + N_*^{-\frac{1}{2}}
\int_0^{2\pi/\beta} dt \, 
u_*^\dagger(t) p(t) u_*(t) N_*^{-\frac{1}{2}}.
\label{phistar}
\eeq

\vskip 1cm
\setcounter{equation}{0}
\makeatletter
\renewcommand\theequation{D\arabic{equation}}
\makeatother

\leftline{\large\bf Appendix D: The SU(3) Case}
\vskip 3mm
\noindent

We first consider the Nahm data for three monopoles which makes a
single instanton on $R^3\times S^1$, or a
caloron~\cite{murray,caloron,kraan}. As shown in Appendix B, the Nahm
equation is defined over three auxiliary time interval, $[t_1,t_2],
[t_2,t_3], [t_3,t_1+\frac{2\pi}{\beta}]$, where $\beta$ is the
circumference of $S^1$.  The Nahm equation is almost trivial and the
Nahm data gives the position vectors of magnetic monopoles as follows:
\beqn
& & {\bf T}_1=-{\bf x}_\alpha = (0,0, R) ,\,\, t\in (t_1,t_2), \nonumber \\
& & {\bf T}_2=-{\bf x}_\beta =(0,0,0) ,\,\, t \in (t_2,t_3), \nonumber \\
& & {\bf T}_3=-{\bf x}_3=(0,0,-K) ,\,\,\, t\in (t_3, t_1+\frac{2\pi}{\beta}),
\eeqn
where ${\bf x}_\alpha$ and ${\bf x}_\beta$ are the positions of
$\balpha$ and $\bbeta$ monopoles, and ${\bf x}_3$ is the position of
the third monopole. For convenience, we put the third monopole at the
$z$ axis and later on take it to infinity by pushing $K\rightarrow
\infty$. The distance between $\balpha$ and $\bbeta$ monopoles are
$R$.  The jumping condition (B16) satisfied by this Nahm data as
follows:
\beqn
& & w^\dagger_1 = \left(\matrix{\sqrt{2(K+R)} \cr 0 \cr}\right), \nonumber \\
& & w^\dagger_2 = \left(\matrix{0 \cr \sqrt{2R} \cr}\right), \nonumber \\
& & w^\dagger_3= \left(\matrix{0 \cr \sqrt{2K} \cr}\right) .
\eeqn
Then  one
can find the $A_i, b\phi$ field configurations by the ADHMN
method, as explored  in detail in Refs,\cite{piljin,caloron}

For  given solutions of the corresponding  ADHMN equation, there
exist a general method to find the solution of the covariant
four-dimensional Laplacian satisfied by the adjoint Higgs field, as
summarized in  Appendix C. For a single caloron as in our case, we
need to find a continuous and periodic function $p(t)$ on
$[t_1,t_1+\frac{2\pi}{\beta}]$,  for a given $q\in
SU(3)$ which commutes with the asymptotic Higgs value ${\bf h}\cdot
{\bf H}$. The differential equation (\ref{pstar}) for the periodic
$p(t)$  in our context is  given by 
\beqn
&& \partial_t^2 p(t) - 2(K+D)(p(t)-q_1)\delta(t-h_1)  \nonumber\\
&& \,\,\;\;\; -2D(p(t)-q_2)\delta(t-h_2)  - 2K(p(t)-q_3)\delta(t-h_3) = 0 ,
\eeqn
where $q=\diag(q_1,q_2,q_3)$ and $q_1+q_2+q_3=0$.
This equation is very simple to solve, especially in the limit where
$K\rightarrow \infty$. 

There are two independent 
$q$ matrices:
\beqn
&& q_T = {\rm diag}(h_1,h_2,h_3), \nonumber \\
&& q_R = {\rm diag} (\mu_2,-\mu_2-\mu_1, \mu_1),
\label{qtr}
\eeqn
where  $\mu_2=h_3-h_2$ and
$\mu_1=h_2-h_1$, so that $\tr q_T\,q_R=0$.
For each $q$, there exists a corresponding $p(t)$. 
Especially in the relevant interval  $t\in [h_1,h_3]$, for
$q_T$, 
\beq
 p_T= t.
\eeq
For $q_R$, 
\beq
 p_R(t) =\left( \matrix{  p_1 (t-h_2) + c,& \,\,\, t\in [h_1,h_2] \cr
              p_2 (t-h_2) +c,& \,\,\, t\in [h_2,h_3]  \cr} \right. ,
\label{prel1}
\eeq
where
\beqn
& & c=h_2+\frac{1}{2R}(p_2-p_1), \nonumber \\
& & p_1 = \frac{\mu_1-\mu_2-2(\mu_1+2\mu_2)\mu_2
R}{\mu_1+\mu_2+2\mu_1\mu_2 R}, \nonumber \\
& & p_2 = \frac{\mu_1-\mu_2+2(2\mu_1+\mu_2)\mu_1
R}{\mu_1+\mu_2+2\mu_1\mu_2 R} .
\eeqn
The $p_T$ can be regarded case where $p_1=p_2=1$.

Following the ADHMN method of the $b\phi$ field closely, as explored
in Ref.~\cite{piljin}, we can solve easily the ADHMN equations (B18)
for a given Nahm data (D1) and (D2).  Especially one can see easily
that the solutions ADHMN equation for the interval $[t_3,
t_1+\frac{2\pi}{\beta}]$ goes to zero like $1/\sqrt{K}$, similar to
the $SU(2)$ case in Ref.~\cite{caloron}. Thus, there will be no
nontrivial contribution from the interval $[t_3,
t_1+\frac{2\pi}{\beta}]$. Then, we can now construct the solution of
the second BPS equation (\ref{gauss2}) by using Eq.~(\ref{phistar}) of
Appendix C.  From Eq.~(\ref{phistar}) and the solution of the ADHMN
equation in Ref.~\cite{piljin}, we can easily construct the $3\times
3$ adjoint Higgs field which satisfies the second BPS equation
(\ref{gauss2}). The solution is
\beq
\Lambda(x)= \left(\matrix{ \phi_{(1)} & \phi_{(3)} \cr
                    \phi_{(3)}^\dagger & \phi_{(2)} \cr} \right),
\label{Lamb}
\eeq
where
\beqn
& & \phi_{(1)} = N^{-\frac{1}{2}}(p_1 K_L+ p_2 K_R )
N^{-\frac{1}{2}} +c I_{2\times 2}, \nonumber \\
& & \phi_{(2)}= 2R L^2  \,\,(0,1)(p_1N_L^{-1} K_L N_L^{-1} + p_2 N_R^{-1} K_R
N_R^{-1}) \left(\matrix{0 \cr 1 \cr}\right)
+ c - \frac{p_2-p_1}{2R} S^\dagger S , \nonumber\\
& & \phi_{(3)} =  N^{-\frac{1}{2}} \left(-p_1K_LN_L^{-1}+p_2 K_R
N_R^{-1}\right)\left( \matrix{0 \cr 1 \cr} \right) \sqrt{2R} L , 
\eeqn
where  ${\bf y}_1={\bf x}-{\bf x}_1$,  ${\bf y}_2={\bf x}-{\bf x}_2$, and 
\beqn
&& N_L = \frac{1}{|{\bf y}_1|} \sinh(\mu_1 y_1) e^{-\mu_1 {\bf y}_1\cdot
\sigma}, \nonumber \\
&& N_R = \frac{1}{|{\bf y}_2|} \sinh(\mu_2 y_2) e^{\mu_2 {\bf y}_2\cdot
\sigma}, \nonumber \\
&& N= N_L + N_R , \nonumber\\
&& K_L = \frac{1}{2y_1}\hat{\bf y}_1\cdot \sigma \bigl[ \mu_1 e^{-2\mu_1
{\bf y}_1\cdot \sigma} - N_L\bigr], \nonumber \\
&& K_R = \frac{1}{2y_2}\hat{\bf y}_2\cdot \sigma \bigl[ \mu_2 e^{2\mu_2
{\bf y}_2\cdot \sigma} - N_R\bigr], \nonumber \\
&& L= \frac{1}{\sqrt{(y_1 \coth \mu_1 y_1+ y_2 \coth \mu_2y_2 )^2
-R^2}}, \nonumber \\
&& S^\dagger S = \frac{y_1 \coth \mu_1 y_1 +y_2\coth \mu_2 y_2 -R}{y_1
\coth \mu_1 y_1 +y_2\coth \mu_2 y_2 +R}.
\eeqn
When $p_1=p_2=1$, we have the solution corresponding to the
$p_T$, which is of course the  original Higgs field, $b\phi$,
itself.

Here only useful part of this explicit solution is its asymptotic form
in the limit where $|{\bf x}| >> R, \mu_1^{-1}, \mu_2^{-1}$. As in
Ref~\cite{piljin},  we can find the asymptotic form of this solution easily.
In the unitary gauge, its asymptotic limit of
Eq.~(\ref{Lamb}) for $q_T$ and $q_R$ of Eq.~(\ref{qtr}) become
Eqs.~(\ref{lamb1}) and (\ref{lamb2}) in
Section 3. 

\vskip 1cm
\setcounter{equation}{0}
\makeatletter
\renewcommand\theequation{E\arabic{equation}}
\makeatother

\leftline{\large\bf Appendix E: Energy Density and Angular Momentum}
\vskip 3mm
\noindent

Here we want to point out that  energy density and total angular
momentum become considerably simpler for the self-dual
configurations. Using the self-dual equations, one can also simplify
the energy density to be 
\beqn
{\cal H}({\bf x}) &=& \tr \left\{ E_i^2 + B_i^2 + (D_0 b\cdot \phi)^2 + (D_i
b\cdot \phi)^2 + (-ie [a\cdot \phi,
b\cdot \phi])^2  \right\} \nonumber \\
&=&  \partial_i^2 \tr [(a\cdot \phi)^2 +(b\cdot \phi)^2] 
\eeqn
where we used the result that $D_0\zeta_I=ie[a\cdot\phi,\zeta_I]=0$.

The most general BPS solutions carry both electric and magnetic
charges and will have nonzero angular momentum in general. The
angular momentum of a BPS configuration is 
\beqn
J^i &=& -2 \int d^3x \,\, \epsilon_{ijk} \, x^j\, \tr \biggl\{ 
\epsilon_{klm} E_l B_m + D_0\phi_I D_k \phi_I \biggr\} \nonumber \\
&=& -2\int d^3x \,\, (x^j \partial_i - \delta^j_ix^l \partial_l)
\,  \tr(a\cdot\phi D_j\,b\cdot\phi )
\eeqn
The angular momentum is a vector quantity and so should depend on the
internal structure of the BPS configuration. While we do not pursue in
the paper, we expect that both energy density and  angular momentum
can be simplified  further.


\newpage




\begin{thebibliography}{99}



\bibitem{oren}
O. Bergman, {\it Three-pronged strings and 1/4 BPS states in N=4
Super-Yang-Mills Theory}, hep-th/9712211

\bibitem{witten} 
E. Witten, Nucl. Phys.  {\bf B460}, 335 (1996), hep-th/9510135.

\bibitem{prong1} %prong ref
O. Aharony, J. Sonnenschein and S. Yankielowicz, Nucl. Phys.  {\bf
B474}, 309  (1996), hep-th/9603009; J.H. Schwarz, {\it Lectures on
Superstring and M-theory dualities}, hep-th/9607201.

\bibitem{prong2}
K. Dasgupta and S. Mukhi, {\it BPS dynamics of triple (p,q) string
junction}, hep-th/9711094; \\
A. Sen, {\it String network}, hep-th/9711130;\\
S.J. Rey and J.T. Yee, {\it BPS dynamics of triple $(p,q)$ string
junction}, hep-th/9711202;\\ 
M. Krogh and S. Lee, {\it String network from M-theory},
hep-th/9712050;\\ 
Y. Matsuo and K. Okuyama,
{\it  BPS condition of string junction from M-theory},
hep-th/9712070.

\bibitem{oren2}
O. Bergman and A. Fayyazuddin, {\it String junctions and BPS states in
Seiberg-Witten theory}, hep-th/9802033; A. Mikahilov, N. Nekrasov and
S. Sethi, {\it Geometric realizations of BPS states in N=2 theories},
hep-th/9803142.

\bibitem{hasimoto}
K. Hasimoto, H. Hata and N. Sasakura, {\it 3-string junction and BPS
saturated solutions in SU(3) supersymmetric Yang-Mills theory},
hep-th/9803127.

\bibitem{fraser}
C. Fraser and T. J. Hollowood, Phys. Lett.  {\bf B402}, 106 (1997),
hep-th/9704011.


\bibitem{erick}
E.J. Weinberg,  Nucl. Phys.  {\bf B167}, 500 (1980).


\bibitem{adhm}
M.F. Atiyah, N.J. Hitchin, V.G. Drinfeld and Yu.I. Mannin,
Phys. Lett. {\bf 185B}, 185 (1978); N.H. Christ, E.J. Weiberg and
N.K. Stanton, Phys. Rev. D {\bf 18}, 2013 (1978); E. Corrigan,
D. Fairlie,  P. Goddard, and S. Templeton, Nucl. Phys.  {\bf B140}, 31
(1978);  E. Corrigan,  P. Goddard, and S. Templeton, {\it ibid}, {\bf
B151}, 93 (1979).



\bibitem{nahm}
W. Nahm, Phys. Lett. {\bf 90B}, 413 (1980); in {\it
Monopoles in quantum field theory}, edited by N. S. Craigie et al.
(World Scientific, Singapore, 1982); in {\it Structural Elements in
Particle Physics and Statistical Mechanics}, edited by J. Honerkamp
et al.  (Plenum, New York, 1983).   


\bibitem{piljin}
E.J. Weinberg and P. Yi, {\it Explicit multimonopole solutions in
$SU(N)$ gauge theory}, hep-th/9803164.

\bibitem{osborn2} % scalar field 
N. Dorey, V.V. Khoze and M.P. Mattis, Phys. Rev. D {\bf 54}, 1921
(1996), hep-th/9603136; H. Osborn, Ann. Phys. (N.Y.) {\bf 135}, 373
(1981). 


\bibitem{murray}
W. Nahm,  in {\it Group Theoretical methods in physics,} edited by
D. Denardo et al. (Springer-Verlag, Berlin, 1984);
H. Garland and M. Murray,  Commun. Math. Phys. {\bf 120}, 335  (1988). 


\bibitem{caloron}
K. Lee and P. Yi, Phys. Rev. D {\bf 56} (1997) 3711, hep-th/9702107; 
K. Lee, {\it Instantons and magnetic monopoles on $R^3\times S^1$ with
arbitrary simple gauge groups}, hep-th/9802012;
K. Lee and C. Lu, {\it SU(2) calorons  and magnetic monopoles}, 
hep-th/9802108.


\bibitem{kraan}
T. C. Kraan and P. van Baal, {\it Exact T-duality between caloron and
Taub-NUT spaces}, hep-th/9802049.


\bibitem{ward}
R. S. Ward, Commun. Math. Phys. {\bf 79}, 317 (1981).


\bibitem{mans}
M. Henningson,  Nucl. Phys.  {\bf B461}, 101 (1996), hep-th/9510138.


\bibitem{manton}
N. Manton and G. Gibbons, Phys. Lett. {\bf B356}, 32 (1995); K. Lee,
E. J. Weinberg, and P. Yi, Phys. Rev. D {\bf 54}, 1633 (1996). 

\bibitem{nonabelian}
K. Lee, E. J. Weinberg, and P. Yi, Phys. Rev. D {\bf 54}, 6351 (1996).

\bibitem{multi}
T. Kawano and K. Okuyama, {\it String network and 1/4 BPS states 
in N=4 SU(n) supersymmetric Yang-Mills theory}, hep-th/9804139.

\bibitem{kol}
O. Bergman and B. Kol, {\it String webs and 1/4 BPS monopoles},
hep-th/9804160.

\end{thebibliography}
\end{document}